\documentclass{mn2e}
\input{epsf}
\newcommand{\be}{\begin{eqnarray}}
\newcommand{\ee}{\end{eqnarray}}

\begin{document}
\title{Semi analytical description of formation of
galaxies and clusters of galaxies}
\author[Demia\'nski \&  Doroshkevich ]
       {M. Demia\'nski$^{1,2}$,  A.G. Doroshkevich$^{3}$,\\
        $1$Institute of Theoretical Physics,
                       University of Warsaw,
                       00-681 Warsaw, Poland\\
        $2$Department of Astronomy, Williams College,
           Williamstown, MA 01267, USA\\
        $3$Astro Space Center of Lebedev Physical
           Institute of  Russian Academy of Sciences,
                        117997 Moscow,  Russia\\
}

\date{Accepted ...,
      Received ...,
        in original form ... .}

%%\begin{document}
\maketitle

\begin{abstract}
We apply the well known semi analytical model of formation of 
DM halos to discuss properties of the relaxed objects
dominated by the DM component (such as the first and dSph 
galaxies and/or clusters of galaxies). This approach allows us
to obtain a simple but more detailed description of evolution 
of the first galaxies. It reveals also links between the
observed characteristics of the relaxed DM halos and the 
initial power spectrum of density perturbations. Results of
our analysis of the observed properties of $\sim 40$ DM 
dominated galaxies and $\sim 100$ clusters of galaxies are 
consistent with the $\Lambda$CDM like power spectrum of 
initial perturbations down to the scale of $\sim 10kpc$. 
For the DM dominated objects the scaling relations are also 
discussed.
\end{abstract}

\section{Introduction}

The formation and evolution of the first galaxies at redshifts
$z\geq 8$ is one of the most interesting problems of modern
cosmology and it is closely connected with many other unresolved
problems. Among others there are the secondary ionization of
the Universe at redshifts $z_{ri}\simeq 10$ implied
by the WMAP observations (Komatsu et al. 2011; Larson et
al. 2011) and recently confirmed by the PLANCK mission (Ade
2013), the formation and evolution of stars with the
primeval chemical composition, the matter enrichment by metals,
evolution of observed galaxies at redshifts $z\geq 6$,
the high redshifts observations of massive galaxies and super
massive black holes and so on (see, e.g., Wiklind et al.
2008; Mancini et al. 2009; Ouchi 2009; Vestergaard 2009; Trenti
et al. 2009, 2010; Kelly 2010; Haiman 2010; Schaerer \&
de Barros 2010; Gonzales et al. 2010, 2012; Shull et al. 2011;
Bouwense et al. 2011; Oesch et al. 2013; Ellis et al. 2013;
Barone--Nugent 2013; Salvadori et al. 2013; Wyithe et al. 2013).

It is specially interesting that observations of the farthest 
quasars and galaxies show that the reionization of the
hydrogen fraction of the intergalactic matter had just been 
completed already at $z\sim 7 - 8$ while ionization of HeII
occured only at $z\sim 3$ (Jakobsen et al. 1994; Hogan et al. 
1997; Smette et al. 2002; Fan et al. 2004, 2006;
Furlanetto and Oh 2008; Trenti et al. 2009; Lehnert et al. 
2010; Robertson et al. 2010). This means that at least at
$z\sim 5 - 7$ the ultraviolet (UV) radiation with energy 
$h\nu\geq 50$ eV is weak and it is mainly generated by 
quasars at $z\leq 4$. The current status of these problems 
is discussed in many recent reviews (see, e.g., Bromm \& Yoshida
2011; Johnson 2011; Kravtsov \& Borgani 2013).

It is very difficult to find or even estimate characteristics 
of the first galaxies formed from matter of the
primordial chemical composition. The absence of metals makes c
ooling of matter and formation of stars  more difficult
and leads to well known peculiarities in the evolution of 
such objects. First of all it is the high typical mass of
first stars ($100 - 1000 ~M_\odot$) and significant energy 
that they eject in the UV region and during the ultimate
explosion as supernovae (see, e.g., Tumlinson et al. 2004; 
Trenti\,\&\,Shull 2010; Bromm \& Yoshida 2011). At the same
time radiation of these stars generates the Lyman -- Werner 
(LW) and infrared (IR) backgrounds that destroy the $H^-$
ions and $H_2$ molecules what delays the cooling of matter 
and slows down the process of star formation.

Some observations (e.g., Bouwens et al. 2011) show that the 
observed rate of star formation and predicted UV radiation
cannot ionize the Universe at $z\sim 10$. Alternative 
explanation assumes that properties of galaxies at small and 
high redshifts can be quite different (Ouchi et al. 2009; 
Gonzales et al. 2010, 2012; Schaerer \& de Barros 2010) and 
that low luminosity galaxies are dominant during the epoch 
of reionization. However this would lead to more efficient
generation of LW and IR backgrounds what in turn would slow 
down formation of the low mass galaxies. Perhaps a more
promising way to alleviate this problem is to take into 
account the non thermal radiation of matter accreted onto 
black holes what changes the spectrum of UV background, 
decreases the LW background and promotes the ionization of 
the Universe. Perhaps this effect can be observed as small 
distortion of the background light caused by emission in 
the He lines such as $\lambda =304A$ and $\lambda=584A$ 
shifted to the redshift of reionization $z_{ri}\sim 10 - 15$. 
More detailed discussion of these problems can be found 
in Meiksin (2009); Trenti et al. (2009, 2010); Shull et al. 
(2011); Giallongo et al. (2012); Barone--Nugent (2013); 
Ceverino et al. 2013).

Numerical simulations provide a powerful method of 
investigation of the epoch of reionization (e.g., Wise \& 
Abel, 2007, 2008; Greif et al. 2008) and makes it possible 
to study the evolution of first galaxies in more details. 
In particular, they allow to investigate the early anisotropic 
stages of halo formation, to trace the process of the halo
virialization, formation of its internal structure and early 
stages of protostar formation. Such analysis can be
performed in a wide range of halo masses and redshifts what 
allows to improve the description of properties of relaxed
halos of galactic scale and to link them with the power 
spectrum of initial perturbations.

However, possibilities of such simulations are strongly 
limited. These simulations are performed within a small box 
with the comoving size $L\sim 0.7 - 1.5 $ Mpc what corresponds 
to the box mass $M_{box}\sim 10^{10} M_\odot$. So small box 
size artificially suppresses the large scale perturbations and 
the formation of more massive objects, what strongly distorts 
the simulated mass function and increases the expected number 
of low mass objects. Small box distorts also the influence of
neighboring objects, the radiation transfer and the feedback of
UV, LW and IR backgrounds. The star formation, their radiation,
explosion and metal production cannot be simulated and are
introduced by hand as independent factors. This list of
limitations can be continued.

It is therefore interesting to consider a more rough but 
simple model of formation and evolution of early galaxies. 
In this paper we propose to use for such analysis the semi 
analytic approach based on the approximate analytical
description of the basic DM structure of collapsed halos 
and numerical estimates of the thermal evolution of the
baryonic component. During the last fifty years similar 
models have been considered and applied to study  various
aspects of nonlinear matter evolution (see, e.g., Peebles 
1967; Zel'dovich \& Novikov 1983; Fillmore and Goldreich
1984; Gurevich \& Zybin 1995; Bryan \& Norman 1998; Lithwick 
\& Dalal 2011).  In this regard it is important to note
that the DM halos are formed before stars appear (see, e.g., 
Kaviraj et al. 2013).

Formation of DM halos is a complex process with strongly
anisotropic matter collapse during both earlier and final periods.
Moreover sometimes this process is interrupted by the violent
merging of neighboring halos. This implies that the simple
spherical model of halos formation can not adequately to describe
the present observational data. However properties of the steady
state virialized DM objects are mainly determined by the
integral characteristics of protoobjects and are only weakly
sensitive to details of their evolution. This is clearly seen
in numerous simulations which show that the Navarro -- Frenk
-- White (NFW) density profile is very stable and is formed
in majority of simulated DM halos.

The same simulations show also that properties of the cores of
virialized DM halos are established during the early period
of halos formation and later on the slow pseudo--evolution of
halos dominates (see, e.g., Diemer et al. 2013). This means
that properties of halo cores only weakly depend on the
halo periphery and are determined mainly by their mass and the
redshift of formation (Klypin et al. 2011). Using these results
we formulate a rough two parametric description of all the basic
properties of DM halos. These parameters are the virial mass of
halos and the redshift of their formation.

First of all this approach allows us to reveal the close
correlation between the redshift of formation and virial
masses of halos with the initial power spectrum of density
perturbations. It also allows to reconsider some of the widely
discussed scaling relations between observed characteristics
of galaxies (see, e.g., Spano et al., 2008; Donato et al. 2009;
Gentile et al. 2009; Hyde \& Bernardi, 2009; Salucci et al.
2011; Mosleh et al. 2011; Besanson 2013). Usually they are
related to properties of luminous matter, such us fundamental
plane, luminosity - velocity dispersion, or mass -- size
relations. Non the less they actually characterize the mass
and entropy profile of halos and its formation in the course
of violent relaxation of the compressed DM component.

Of course this approach is applied for the DM dominated halos
only as the dissipative evolution of the baryonic component
distorts properties of the cores of DM halos. In spite of this
we can use this model in three ways:
\begin{enumerate}
\item{} The density and temperature profiles of
the DM component can be considered as the initial conditions for
numerical analysis of the process of cooling and compression of
the baryonic clouds within the stable DM halos. In particular,
with this approach it was possible to estimate the evolution of
the Jeans mass of cold baryons down to the masses of Pop. III
stars.
\item{} We can use the redshift of the DM halos formation
as a parameter that characterizes the 'frozen' properties of
the central region of DM halos. This redshift correlates
with the virial mass of halos and so with the initial power
spectrum. Thus this approach allows us to reveal the impact
of initial conditions on the observed characteristics of the
DM dominated objects.
\item{}This approach allows us to clarify also some of the
widely discussed scaling relations that are applied to the
DM dominated objects.
\end{enumerate}

However potential of such approach should not be overestimated.
Thus clusters of galaxies presented in Pratt et al.
(2009) illustrate large scatter of matter distribution in
the observed DM halos.

This paper is organized as follows. In Sec. 2 the basic
relations and assumptions of our approach are formulated and the
expected properties of the DM halos are presented. Properties
and evolution of the baryonic component are described in Sec. 3.
Mass dependence of the redshift of formation and the scaling
relations for observed object are considered in Sec. 4.
Discussion and conclusions can be found in Sec. 5.

\subsection{Cosmological parameters}

In this paper we consider the spatially flat $\Lambda$CDM model
of the Universe with the Hubble parameter, $H(z)$, the critical
density $\rho_{cr}$, the density of non relativistic matter,
$\langle\rho_m(z)\rangle$,  and the mean density and mean number
density of baryons, $\langle\rho_b(z)\rangle\,\&\,\langle n_b(z)
\rangle$, given by:
\[
H^{2}(z) = H_0^2[\Omega_m(1+z)^3+\Omega_\Lambda],\quad H_0=
100h\,{\rm km/s/Mpc}\,,
\]
\[
\langle\rho_m(z)\rangle =
2.5\cdot 10^{-27}z_{10}^3\Theta_m\frac{g}{cm^3}=
3.4\cdot 10^{4}z_{10}^3\Theta_m\frac{M_\odot}{kpc^3} \,,
\]
\be
\langle\rho_b(z)\rangle = {3H_0^2\over 8\pi G}\Omega_b(1+z)^3
\approx 4\cdot 10^{-28}z_{10}^3\Theta_b\frac{g}{cm^3} \,,
\label{basic}
\ee
\[
\rho_{cr}=\frac{3H^2}{8\pi G},\quad z_{10}=\frac{1+z}{10},
\quad \Theta_m=\frac{\Omega_mh^2}{0.12},\quad \Theta_b=
\frac{\Omega_bh^2}{0.02}\,.
\]
Here $\Omega_m=0.24\,\&\,\Omega_\Lambda=0.76$ are dimensionless
density of non relativistic matter and dark energy, $\Omega_b
\approx 0.04$ and $h=0.7$ are the dimensionless mean density of
baryons, and the Hubble constant measured at the present epoch.
Cosmological parameters presented in recent publication of the
PLANCK collaboration (Ade et al. 2013) slightly differ from
those used above (\ref{basic}).

For the $\Lambda$CDM cosmological model the evolution of
perturbations can be described with sufficient precision by
the expression
\be
\delta\rho/\rho\propto B(z),\quad B(z)^{-3}\approx \frac{1-
\Omega_m+2.2\Omega_m(1+z)^3}{1+1.2\Omega_m}
\label{Bz}
\ee
(Demianski, Doroshkevich, 1999, 2004; Demianski et al. 2011) 
and for $\Omega_m\approx 0.25$ we get
\be
B^{-1}(z)\approx\frac{1+z}{1.35}[1+1.44/(1+z)^3]^{1/3}\,,
\label{bbz}
\ee
For $z=0$ we have $B=1$ and for $z\geq 1$ B(z) is reproducing
the exact value with accuracy better than 90\%.

For $z\gg 1$ these relations simplify. Thus, for the Hubble
constant and the function $B(z)$ we get
\be
H^{-1}(z)\approx \frac{2.7\cdot
10^{16}}{\sqrt{\Theta_m}}s \left[\frac{10}{1+z}\right]^{3/2},
\quad B(z)\approx \frac{1.35}{1+z}\,,
\label{bzz}
\ee

\section{Physical model of halos formation}

As is commonly accepted in the course of complex nonlinear
condensation the DM forms stable virialized halos with more or
less standard density profile. Numerical simulations show that
after short period of rapid evolution the structure of the
virialized DM halos is quite well described by the spherical
model with the Navarro -- Frenk -- White (NFW) density profile
(Navarro et al. 1995, 1996, 1997; Ludlow et al., 2013). The
basic parameters of the model --  the virial mass, $M_{vir}$,
central density, $\rho_c$, core scale $r_s$, and concentration,
$c$, -- were fitted in a wide range of redshifts and halo
masses in many papers (see, e.g. Klypin et al. 2011).

After the completion of active phase of halo formation its
parameters only weakly depend on the redshift and at
large radius a steeper asymptotic density profile is formed,
\[
\rho(r)\propto r^{-4}\,,
\]
(see, e.g., discussion in Visbal, Loeb,\,\&\,Hernquist, 2012).
But as before the central regions of halos are described by the
NFW profile. For example such virialized objects are observed as
isolated galaxies with the rotation curve $v_c\propto r^{-1/2}$
and/or as high density galaxies within clusters of galaxies,
filaments or other elements of the Large Scale Structure of the
Universe.

These results can be successfully used to roughly estimate  the
mean density and temperature of early galaxies what in
turn allows us to concentrate main attention on the thermal
evolution of the compressed gas. In this way we can consider
the evolution of baryonic component and the formation of the
first stars for a wide range of redshifts and virial
masses of DM halos. Evidently similar approach can be applied
also for investigations of more complex evolution of
halos with the metal enriched baryonic component.
This approach allows us also to estimate the
redshift when the observed DM dominated objects such us the
dSph galaxies and clusters of galaxies were formed.

It is important that the standard description of both the
observed and simulated DM halos links the virial mass and
radius of halos by the condition
\be
M_{vir}=4\pi/3R_{vir}^3\Delta_v\langle\rho_{cr}(z_{cr})\rangle\,,
\label{nfw-vir}
\ee
where $z_{cr}$ is the redshift of relaxation of DM halo,
$\Delta_v=18\pi^2\approx 200$ and $\langle\rho_{cr}(z_{cr})
\rangle$ is the critical density of the Universe at this
redshift. This relation can be applied for halos embedded in a
homogeneous medium -- early galaxies and clusters of galaxies
-- and the value of $\Delta_v$ was derived from the simple model
of spherical collapse that ignores the influence of complex
anisotropic halos environment (see, e.g., Bryan \& Norman 1998;
Vikhlinin et al. 2009; Lloyd--Davies et al. 2011).

Of course, this approach has only limited predictive power.
Thus, it ignores the complex anisotropic matter compression
within filaments and walls before formation of compact halos,
it ignores the effects produced by mergers and so on. These
restrictions do not allow us to consider the process of
generation of the angular momentum of the compressed matter
and to link the properties of DM halos and the rate of star
formation with the primordial characteristics of collapsed
matter such as the anisotropic shape of density peaks and
their environment, the internal velocity dispersion and so
on. Non the less with this approach further progress in the
description and understanding of the complex processes of
formation of DM halos and early galaxies can be achieved.

\subsection{Internal structure of DM halos}

Further on we consider the virialized spherical DM halos
characterized by the virial mass $M_{vir}=10^9M_9M_\odot$ at
the (conventional) redshift of formation $z=z_{cr}$. For any
model the universal mass, $M$, and density, $\rho$, profiles
of virialized DM halo can be taken as follows:
\be
M(x)=M_cf_m(x),\quad \rho(x)=\rho_cf_\rho(x),\quad M_c=
4\pi\rho_c r_s^3\,.
\label{mdens}
\ee
Here $x=r/r_s$ and $r_s(M_{vir},z_{cr}),\,\&\,\rho_c(M_{vir},
z_{cr})$ are the typical size and density of the halos cores. 
For the popular NFW  model the density and mass profiles are
\be
f_\rho=x^{-1}(1+x)^{-2},\quad f_m=ln(1+x)-\frac{x}{1+x}\,,
\label{fmass}
\ee
and $f_m(5)\simeq 1$. For another popular model (Burkert, 1995) 
the density profile is
\[
f_\rho=(1+x)^{-1}(1+x^2)^{-1},\,
\]
\be
f_m=ln[(1+x)\sqrt{1+x^2}]-tg^{-1}(x),\quad f_m(5)\simeq 1.8\,.
\label{bmass}
\ee
These models can be used for $x\leq 5 - 6$. From these 
expressions it follows that at $x\geq 1$ differences between 
these models are quite moderate and our results obtained 
below for the NFW model can be applied with small corrections 
also for the Burkert model. More detailed discussion of these 
models can be found in Penarrubia et al. (2010).

Both the DM and baryonic components are treated as the ideal
gas with the pressure, $P$, temperature, $T$, and the entropy
function, $S$, linked by the usual expressions for the non
relativistic particles:
\be
 P(x)=n(x)T(x)=S(x)n^{5/3}(x),
\label{functions}
\ee
\[
P(x)=P_cf_p(x),\quad T(x)=T_cf_T(x),\quad S(x)=S_cf_s(x)\,,
\]
where the typical temperature, $T_c(M_9,z_{cr})$, entropy,
$S_c(M_9,z_{cr})$, pressure, $P_c(M_9,z_{cr})$ and the number
density of the DM component, $n_{DM}(M_9,z_{cr})f_\rho(x)$,
or baryons, $n_b(M_9,z_{cr})f_\rho(x)$, also depend upon the
virial halo mass $M_9$ and its redshift of formation, $z_{cr}$.

Random variations of the profile and amplitude of the initial
velocity, the initial shape of collapsed clouds, properties of
outer regions of halos and so on lead to random variations
of halos density, shape, profile and other parameters relative
to the accepted mean characteristics. The analysis of available
simulations shows that the probability distribution functions
(PDFs) of these variations are often close to the exponential
ones and therefore their random scatter is close to the mean
values (see, e.g., Press\,\&\,Rybicki, 1993; Demianski et al.
2011). These random variations can change the real parameters
by a factor of 2 - 3.

\subsection{Simple model of early galaxies}

 In this paper we consider a simple physical model of formation
of galaxies based on the following assumptions:
\begin{enumerate}
\item{} We assume that at redshift $z=z_{cr}$ the evolution of
DM perturbations results in the formation of spherical virialized
DM halos with mass $M_{vir}=M_9\cdot 10^9M_\odot$ and the density
profile (\ref{fmass}).
\item{} We do not discuss the dynamics of DM halos formation and
evolution which are accompanied by the progressive matter accretion,
the growth of the halos masses and corresponding variations of
other halos parameters. The real process of halos formation is
extended in time what causes some ambiguity in their parameters
such as the halos masses and the redshift of their formation (see,
e.g. discussion in Diemand, Kuhlen \& Madau 2007; Kravtsov \&
Borgani 2012). In the proposed model the redshift of halo
formation, $z_{cr}$, is identified with the redshift of collapse
of the homogeneous spherical cloud with the virial mass $M_{vir}$,
\be
1+z_{cr}\approx 0.63(1+z_{tr})\,,
\label{zform}
\ee
where $z_{tr}$ is the redshift corresponding to the turn
around moment of the cloud evolution (see discussion in Umemura,
Loeb\,\&\,Turner 1993).
\item{}We assume that in the course of DM halo formation
the main fraction of the baryonic component is heated by the
accompanied shock waves up to the temperature and pressure
comparable with the virialized values of the DM component. 
These proceses provide the formation of equilibrium distribution 
for the baryonic component. 
\item{} We assume that some part of the compressed baryons
is disrupted into a system of subclouds which are rapidly
cooled and transformed into high density starlike subclouds.
Thus, the virialized halo configuration is composed of the
DM particles, the adiabatically compressed hot low density
baryonic gas, and cold high density baryonic subclouds.
\item{} Following Hutchings (2002) we assume that the cooling
of the high density baryonic subclouds with both atomic
($H\,\&\,He$) and molecular ($H_2$) coolants proceeds under the
condition that $P\sim const$. Such cooling of the low mass
subclouds corresponds to the isobaric mode of the thermal
instability.
\item{} We assume that the subclouds with masses larger than
the Jeans mass $M_J(n,T)$ rapidly form first stars
with masses $100 - 1000 M_\odot$ what transforms the DM halos
into the early galaxies.
\end{enumerate}

The evolution of the cooled low mass subclouds can be very
complex. It can be approximated by the isobaric mode of the
thermal instability and therefore it does not preserve the
compact shape of the cooled subclouds. As was discussed in
Doroshkevich and Zel'dovich (1981) the motion of such subclouds
within the hot gas leads to their deformation and less massive
subclouds could be even dissipated. The complex aspherical
shape of such subclouds makes their survival problematic and
requires very detailed investigation to predict their evolution.
It is very difficult to estimate also the efficiency of
transformation of protostars into stars. These problems are
beyond the scope of this paper.

In central regions of halos the gas pressure is supported by
adiabatic inflow of high entropy gas from outer regions of halos
what leads to progressive concentration of baryonic component
within central regions of halos and to formation of massive
baryonic cores (see, e.g. Wise\,\&\,Abel 2008; Pratt et al. 2009).
The structure and evolution of such halos resemble the internal
structure of the observed and simulated galaxies and cluster of
galaxies (see, e.g., Pratt et al. 2009, 2010; Arnaud et al. 2010;
Kravtsov \& Borgani 2012).

\subsection{Mean characteristics of the DM halos}

In this section we consider evolution of the central regions
of virialized DM halos using the NFW approximations
presented in Klypin et al. (2011), Prada et al. (2011), Angulo
et al. (2012). For such halos of galactic scale and at $z\gg 1$
the matter concentration $c(M_9,z_{cr})$ is described by the
expression
\be
c=R_{vir}/r_s\approx 4M_9^{1/6}z_f^{7/3}\delta^{1/3}_r
\epsilon,\quad z_f=(1+z_{cr})/10\,,
\label{nfw-c}
\ee
\[
\epsilon(M,z_{cr})\approx 1+0.31M_9^{-1/4}z_f^{-4},
\quad M_9=M_{vir}/10^9M_\odot\,.
\]
Here $R_{vir}$ is the virial radius of halo
and the factor $\delta_r^{1/3}\sim 0.6 - 1.4$ characterizes the
random scatter of the matter concentration caused by the random
variations of characteristics of the outer regions of halos. It
is important that here we consider expected properties of early
galaxies for which $M_9z_f^{16}\geq 1,\,\epsilon \sim 1$. The
usually discussed model of clusters of galaxies with $\epsilon\gg
1$ is considered below in Section 4\,.

Using the standard relations (\ref{nfw-vir},\,\ref{mdens})
\[
M_{vir}=4\pi \rho_cr_s^3f_m(c)=4\pi/3R_{vir}^3\langle\rho_{vir}
\rangle,\quad \langle\rho_{vir}\rangle=\Delta_v\langle\rho_m
\rangle\,,
\]
we find the virial and core sizes of the halo,
\be
R_{vir}\approx 3.34 \frac{M_9^{1/3}}{z_f} kpc,\quad
r_s=\frac{R_{vir}}{c}=\frac{0.8\,M_9^{1/6}}{z_f^{10/3}
\delta_r^{1/3}\epsilon}kpc\,.
\label{size}
\ee
For the central density of the DM matter, $\rho_c$, and its
number density, $n_{DM}$, we get
\be
\rho_c(M,z)\approx
\frac{\langle\rho_m\rangle\Delta_{v}c^3}{3f_m(c)}\approx
1.3\cdot 10^8z_f^{10}M_9^{1/2}\Theta_\rho\,
\frac{M_\odot}{kpc^3}\,,
\label{dnfw}
\ee
\[
n_{DM}=\frac{\rho_c}{m_{DM}}=6z_f^{10}M_9^{1/2}
\Theta_\rho \frac{m_b}{m_{DM}}cm^{-3},\quad
\Theta_\rho=\frac{\delta_r\epsilon^3}{f_m(c)}\,.
\]
Here $\delta_r\sim 0.1 - 3$ characterizes the discussed above
random variations of the halo parameters relative to the mean
characteristics presented by (\ref{dnfw}).

As was noted in the introduction the important characteristic of
DM halos is the central surface density of the DM component,
$\mu_{cs}$, (Donato et al. 2009; Salucci et al. 2011)
\be
\mu_{cs}=r_s\rho_c\approx 10^2(M_9z_f^{10})^{2/3}\Theta_\mu
\,M_\odot/pc^2,
\label{surf}
\ee
with $\Theta_\mu=\delta_r^{2/3}\epsilon^2/f_m(c)$. The similar
virial surface density,
\be
\mu_{vir}=R_{vir}\langle\rho_{vir}\rangle=22.7M_9^{1/3}z_f^2
\Theta_m\, M_\odot/pc^2\,,
\label{vsurf}
\ee
is close in some respects with such phenomenological concepts
as the Fundamental Plane or mass -- size relation which
also are discussed in many publications (see, e.g., Hyde \&
Bernardi, 2009; Mosleh et al. 2011; Besanson 2013).

The velocity dispersion, $\sigma_v^2(r)$, and the temperature,
$T_{DM}(r)$, within the relaxed DM halo with the NFW
density profile are closely linked to the circular velocity of
DM halos, $v_c^2(r)$,
\[
v_c^2(r)=\frac{GM(r)}{r}=\sigma_0^2\frac{f_m(x)}{x},\quad
\sigma_v^2(x)\approx \frac{v_c^2(x)}{2\sqrt{x}}=
\sigma_0^2\frac{f_m(x)}{2x^{3/2}}\,,
\]
\[
\sigma_0^2\approx 4.5\cdot 10^3M_9^{5/6}z_f^{10/3}\Theta_T\,
km^2/s^2,\quad \Theta_T=\epsilon\delta_r^{1/3}/f_m(c)\,,
\]
where again $x=r/r_s$.
Thus for the temperature of DM component with Maxwellian
velocity distribution we have
\[
T_{DM}=m_{DM}\sigma_v^2/3=T_cf_T(x),\quad
f_T(x)=f_m(x)/x^{3/2}\,,
\]
\be
T_c\approx m_{DM}\sigma_0^2/6\approx 1.1\cdot 10^5M_9^{5/6}
z_f^{10/3}m_{DM}/m_b\Theta_T\,K\,.
\label{tnfw}
\ee
For the pressure of the DM component we get
\[
P_{DM}=n_{DM}T_{DM}=P_cf_p(x),\quad f_p(x)=f_T(x)f_\rho(x)\,,
\]
\be
P_c\approx 0.4(M_9z_f^{10})^{4/3}\Theta_p keV/cm^3,
\quad \Theta_p=\Theta_\rho\Theta_T\,.
\label{pnfw}
\ee
Here $m_{DM}$ and $m_b$ are the masses of DM particles and
baryons. For the NFW model the pressure at the center of halo
is divergent, $f_p(x)\propto x^{-1/2}$ what is another
manifestation of the known core -- cusp problem. This artificial
divergence does not prevent the use of estimates (\ref{tnfw})
and (\ref{pnfw}) in our further discussions.

As is seen from this analysis the NFW halos are a two
parametric sample and all the mean halos properties are
determined by the redshift of halos formation, $z_f=(1+z_{cr})
/10$, and their virial masses $M_{vir}$. This is the direct
result of the fixed density profile (\ref{fmass}) and the
expression for the matter concentration (\ref{nfw-c}).

For the Burkert model we have no quantitative fits for the
halo concentration. However comparing  (\ref{fmass}) and 
(\ref{bmass}) we can expect that the variations of the 
central density profile only weakly influence the 
characteristics of the matter distribution in halos at 
$x\geq 1$ and thus to obtain typical characteristics for 
any halos we can use the relations (\ref{nfw-c})--
(\ref{pnfw}). In this case for Burkert profile the function 
$f_p(x)$ at $x\leq 1$ is similar to the isothermal one,
\be
f_p(x)=f_\rho(x)f_m(x)/x^{3/2}\simeq p_0-p_1x^{3/2}+...\,,
\label{pt_br}
\ee
what eliminates the divergence of pressure at $x=0$ in
(\ref{pnfw}). At $x\geq 1$ the function $f_p(x)$ is close 
to that obtained in (\ref{pnfw}) for the NFW model.

This means that at $x\geq 1$ differences between results
obtained for various density profiles are in the range of
random scatter. Results obtained for the generalized NFW
model (Nagai, Kravtsov, \& Vikhlinin 2007) confirm that
numerical results are only weakly sensitive to moderate
variations of the pressure and density profiles.

\section{Evolution of the baryonic component}

In the course of formation of galaxies the evolution of the
baryonic component is driven by the evolution of the dominant
DM component. In particular, the pressures of compressed DM
and baryonic components are the same. However other
properties of these components differ in many respects. Thus
some fraction of baryons is compressed adiabatically, while
other fraction is compressed by shock waves. Both fractions
are compressed up to the pressure $P_{DM}(x)$ but their
density, temperature and further evolution are different.
Thus the shock compression is unstable and is accompanied
by gas disruption into numerous subclouds. The low mass
fraction of such subclouds can be transformed into
Population III stars. Hot low density baryons form gaseous
component of halos.

\subsection{Adiabatically compressed baryonic component}

In virialized halos the pressures of baryonic and DM components
are equal to each other, $P_b(x)=P_{DM}(x)$. Therefore
for a given $P_b(x)$ properties of the adiabatically compressed
gas depend upon the relic entropy of baryonic
component, $S_{rel}$. This entropy can be determined from the
condition that at $z\sim 100 - 300$ the temperatures of
baryons and relic radiation are close to each other. For
example at $z=100$ we have $T_b(100)\approx 2.7\cdot
10^2\approx 0.023eV$, $\langle n_b(100)\rangle\approx
0.24cm^{-3}$ and the relic entropy of baryons is
\be
S_{rel}\approx S_0=T_b(z)\langle n_b(z)\rangle^{-2/3}\approx
0.06 eV cm^{2}\frac{100}{1+z}\,,
\label{rel}
\ee
At the same time the relic (frozen) concentration of electrons
and protons at $z\leq 100$ is small, $f_e=f_p\approx 10^{-4}$
what decelerates the formation of $H_2$ molecules and prevents
cooling of the baryonic component below the  temperature $T_b(x)
\leq 10^4K$. This means that for this component
\[
P_b(x)=P_{DM}(x),\quad T_b=T_{bc}f_P^{2/5}(x),
\quad n_b=n_{bc}f_P^{3/5}(x)\,,
\]
where the function $f_P(x)$ is determined by (\ref{pnfw},
\ref{pt_br}) and
\be
T_{bc}=S_{rel}(P_c/S_{rel})^{2/5}\approx 2 (M_9z_f^{10})^{8/15}
\Theta_{bc}eV\,,
\label{t-lim}
\ee
\[
n_{bc}=(P_c/S_{rel})^{3/5}\approx 2\cdot 10^2(M_9
z_f^{10})^{4/5}\Theta_p/\Theta_{bc}cm^{-3}\,.
\]
\[
\Theta_{bc}=\epsilon^{8/5}\delta_r^{8/15}f_m^{-4/5}(c)(S_{rel}
/S_0)^{3/5}\,.
\]
If the mass and the redshift of formation of the clouds of gas
are limited by the condition
\[
T_b(x)\leq 10^4K\,,
\]
 or by the corresponding restriction for the pressure
\be
P\leq 50eV/cm^3(T_b/10^4K)^{5/2}(S_0/S_{rel})^{3/2}\,,
\label{limit}
\ee
\[
M_9z_f^{10}\leq 0.2\Theta_p^{-3/4}(T_b/10^4K)^2(S_0/S_{rel})\,,
\]
then the cloud evolves in the adiabatic regime and
the Jeans mass of the collapsed clouds $M_J$ remains large
\be
S_{bar}\approx S_{rel},\quad M_J\approx 2.4\cdot 10^5M_\odot
(1 cm^{-3}/n_{bc})^{1/2}\,.
\label{adiabat}
\ee
In this case formation of the real stars is strongly suppressed.

However, if the condition (\ref{limit}) is violated and the
collapsed gas is heated up to temperature $T_b(x)\geq 10^4K$ then
the concentrations of both electrons and $H_2$ molecules  are
rapidly increasing, gas is cooled and forms the high density
low mass baryonic subclouds. The same process rapidly occurs
also when the matter ionization is caused by external sources of
the UV radiation.

\subsection{Formation of high density baryonic subclouds}

Some fraction of the baryonic component is compressed and heated
in the shock waves generated in the course of the matter
infall into the DM potential well. For this fraction the
pressure, velocities and temperature of both baryonic and
DM components are quite similar to each other and are given
by expressions (\ref{tnfw},\,\ref{pnfw}):
\be
P_b(x)=P_{DM}(x),\quad T_b(x)\approx T_{DM}(x)m_b/m_{DM}\,.
\label{shock}
\ee
In this case the density and entropy of the shock compressed
gas are
\be
n_b(x)\approx P_{b}(x)/T_{b}(x)\approx 44z_f^{10}M_9^{1/2}
\Theta_\rho f_\rho(x)\,cm^{-3}\,,
\label{nfw-ent}
\ee
\[
S_{bar}(x)=P_b/n_b^{5/3}\geq S_{rel}\,,
\]
where the function $f_\rho(x)$ and the parameter
$\Theta_\rho$ were defined above.

Instability of the shock compression and heating of baryons
and their subsequent cooling lead to their fragmentation and
formation of subclouds with high $n_b$ and small $T_b$. If the
formation of DM halo occurs with the cosmological characteristic
time,
\[
t_{cosm}=H^{-1}(z)\approx 2.7\cdot 10^{16}z_f^{-3/2}s\,,
\]
then the characteristic hydrodynamical time for the shock
compressed baryonic component at $z_f\geq 1$ is much smaller,
\be
\tau_{hyd}=\frac{1}{\sqrt{4\pi G\rho_b}}\sim \frac{1.4\cdot
10^{14}s}{M_9^{1/4}z_{f}^{5}\Theta_\rho^{1/2}}f_\rho^{-1/2}
(x)\ll H(z)^{-1}\,,
\label{thyd}
\ee
and cooling of the highly ionized compressed gas occurs
even more rapidly. For instance for the free -- free cooling
the characteristic time, $\tau_{ff}$, is
\be
\tau_{ff}\simeq \frac{5\cdot 10^{12}s}{z_f^{8}M_9^{1/2}}
\,\frac{\Theta_T^{1/2}}{\Theta_\rho}\frac{f_T^{1/2}(x)}{f_\rho(x)}
\ll\tau_{hyd}\,.
\label{ff}
\ee
The impact of other atomic processes (recombination and excitation
of H and He etc.) significantly decreases even this
characteristic time. The characteristic time for the hydrogen
recombination is also quite small,
\[
\tau_{rec}\approx 3.1\cdot 10^{11}s \,z_f^{-8.4}
M_9^{0.06}\Theta_\rho f_\rho(x)/\sqrt{\Theta_Tf_T(x)}
\ll \tau_{hyd}\,,
\]
what ensures almost equilibrium ionization of baryons
throughout the period of thermal evolution for $T_b\geq 10^4K$.

However for temperature $T_b\leq 10^4K$ the cooling process
determined by the molecular hydrogen is slower and the
characteristic time of gas cooling becomes comparable with
$\tau_{hyd}$ (\ref{thyd}). This means that the gas pressure
within cold subclouds is retained near $P_b(x)$ (\ref{shock})
and the cooling is accompanied by a corresponding growth of
baryonic density.

Owing to the thermal instability such evolution strongly favors
further fragmentation of the cooled gas. The gravitationally
bounded cold high density subclouds with masses larger than
the Jeans mass, $M\geq M_J$, could be transformed into stars.
The formation of stars with the mass $M_{st}$ is regulated
by the drop of temperature, $T_4=T_b/10^4K$,
\be
M_{st}\geq M_J\approx 2\cdot 10^7M_\odot\frac{T_4^{3/2}}{
n_3^{1/2}}=\frac{2.5\cdot 10^6M_\odot T_4^2}{(M_9z_f^{10})^{2/3}}
\sqrt{\frac{5}{\Theta_\rho}}\,,
\label{jeans}
\ee
where $n_3=n_b/1cm^{-3}$ is the number density of cooling gas.
Some part of the gas is concentrated near the center of the
host DM halo (Wise\,\&\,Abel 2008; Pratt et al. 2009 2010)
forming the baryonic core.

\subsection{Numerical estimates}

The cooling process of the baryonic component can be followed
numerically by solving the equations of thermal balance and
evolution of nine components, namely, the electrons,
$e$, protons, $p$, neutral and molecular hydrogen, $H\,\&\,
H_2$, ions $H^-\,\&\,H_2^+$, and neutral and ionized helium,
$He, He^+, He^{++}$. The kinetic coefficients and the cooling
rates used here are taken from Hutchings et al. (2002).

As is well known there are two different regimes of cooling of
the compressed gas. Thus, very slow cooling takes place
when temperature of the compressed gas (\ref{tnfw},\,
\ref{shock}) does not exceed $\sim 10^4K$. In this case
the low electron concentration $y_e=n_e/n_{b}\sim 10^{-4}$
created at high redshifts $z\sim 100$ remains unchanged, the
formation of molecules $H_2$ and gas cooling are slow and
formation of starlike subclouds is strongly delayed. On
the other hand when temperature of the compressed gas
(\ref{tnfw},\,\ref{shock}) exceeds $10^4K$ and the strong
ionization of hydrogen takes place then the concentrations of
both electrons and molecules $H_2$ strongly increase, the
compressed gas rapidly cools down to temperature $T_b\simeq
100K$ and forms starlike high density subclouds.

\begin{figure}
\centering
\epsfxsize=8. cm
\epsfbox{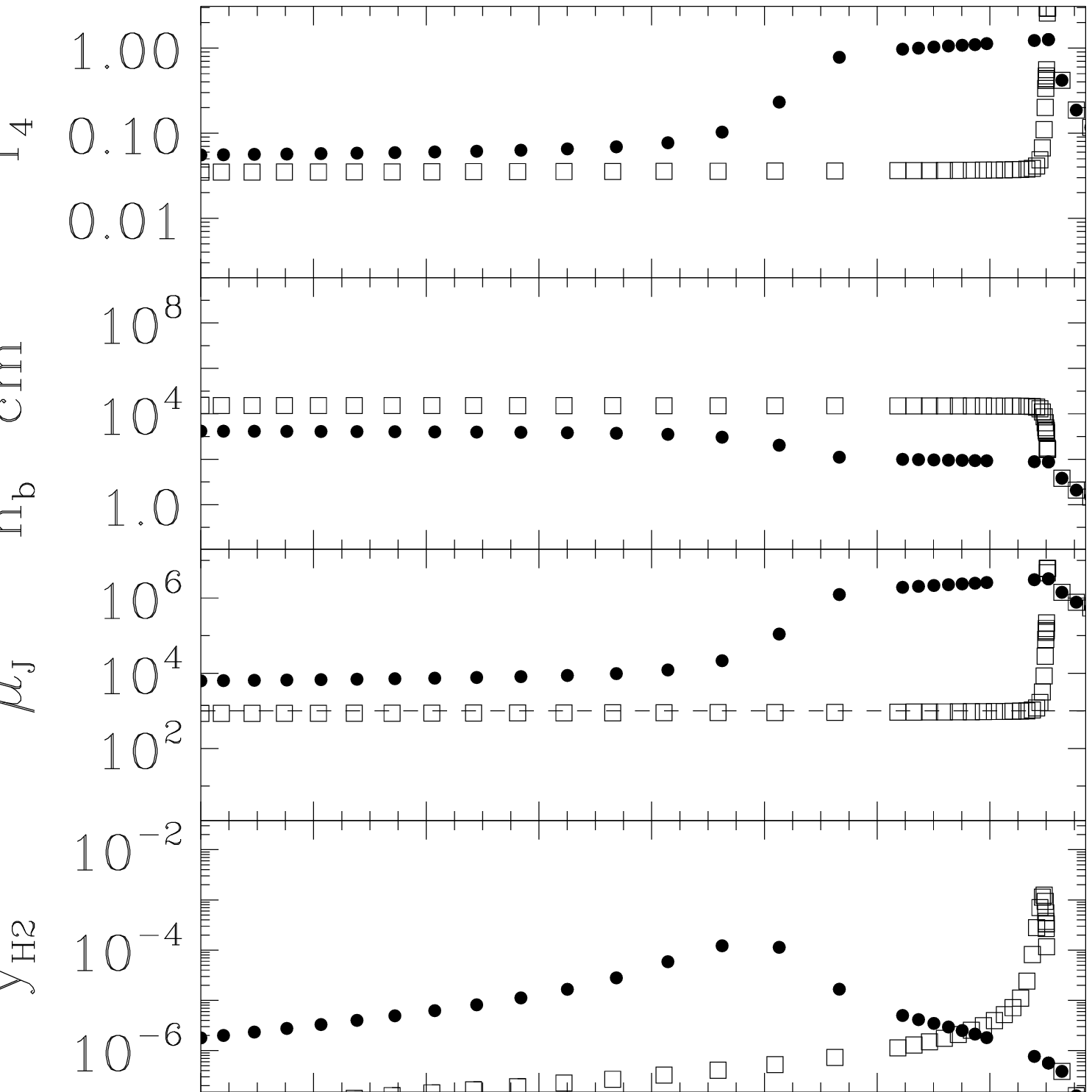}
%%\epsfbox{/home/dorr/REC1/THR/H2therm/figs/tmp_652.eps}
\vspace{1.cm}
\caption{Redshift variations of the temperature, $T_4$,
baryonic number density, $n_b$, Jeans mass, $\mu_J=M_J/M_\odot$,
and $H_2$ concentration, $y_{H2}$, within halos formed at
$z_{cr}=25,\,z_f=2.5$, with the virial masses $M_{vir}\approx
5\cdot 10^5M_\odot$ (points) and $M_{vir}\approx 9\cdot 10^5
M_\odot$ (squares).
}
\label{temp310}
\end{figure}

As is seen from (\ref{tnfw}, \ref{shock}) temperature of
the compressed baryons is a two parametric function.
Thus for a given virial mass $M_9$ it rapidly decreases with
$z_{cr}$ and for some $z_{cr}$ the threshold temperature
$T_b\approx 10^4K$ is reached. This means that at such $z_{cr}$
the baryon cooling and the star formation process are
strongly decelerated. In turn, for a given $z_{cr}$ and halos
of low virial masses the threshold temperature $T_b\approx
10^4K$ cannot be reached. For such virial mass the star
formation process becomes also suppressed. This means that star
formation in low mass halos occurs mainly at higher redshifts
while at $1+z_{cr}\sim 10$ stars cannot be formed within halos
with $M_9\leq 1$.

\begin{figure}
\centering
\epsfxsize=8. cm
\epsfbox{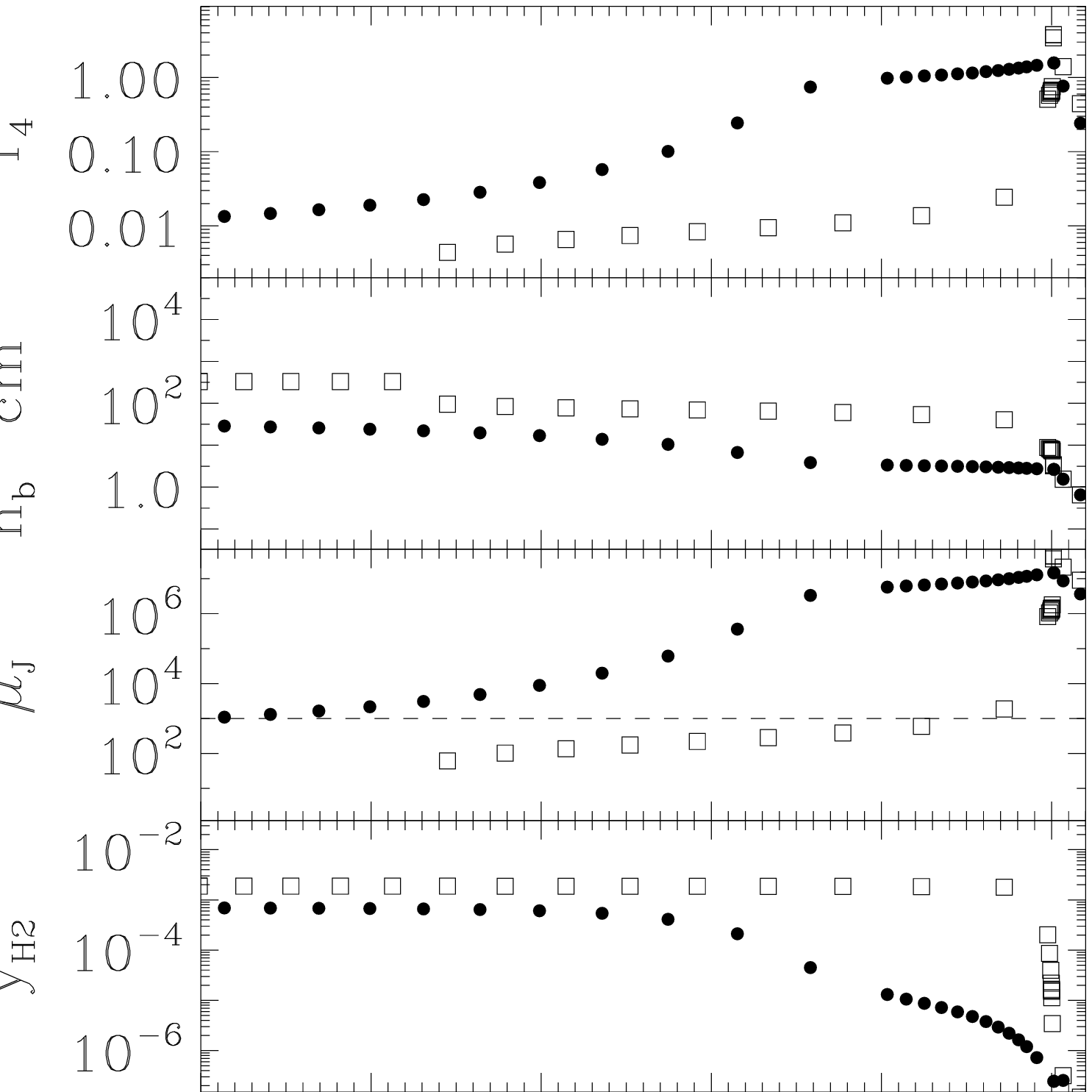}
%%\epsfbox{/home/dorr/REC1/THR/H2therm/figs/temp_932.eps}
\vspace{1.cm}
\caption{Redshift variations of the temperature, $T_4$,
baryonic number density, $n_b$, Jeans mass, $\mu_J=M_J/M_\odot$,
and $H_2$ concentration, $y_{H2}$, within halos formed at
$1+z_{cr}=10,\,z_f=1.0$, with the virial masses $M_{vir}
\approx 10^9M_\odot$ (squares) and $M_{vir}\approx 0.4\cdot
10^9M_\odot$ (points).
}
\label{temp191}
\end{figure}

These statements are illustrated in Figs. \ref{temp310} and
\ref{temp191} where the thermal evolution of compressed gas is
presented for two sets of halo masses and two redshifts of halo
formation. As is seen in both Figures for less massive halos
the rapid atomic cooling at $T_4\geq 1$ is replaced by slower
cooling with $H_2$ molecules at $T_4\leq 1$. In contrast for
more massive halos formed at the same redshift $z_f$ the
cooling rate with $H_2$ molecules remains quite rapid.

For two halos with virial mass $M_{vir}=5\cdot 10^5M_\odot$
and $M_{vir}= 9\cdot 10^5 M_\odot$ formed at $z_{cr}=25$ the
evolution of high density gaseous subclouds is presented in
Fig. \ref{temp310}. The formation of gravitationally bounded
subclouds is restricted by the Jeans mass, $\mu_J(z)$, which
drops down to starlike value $M_J\simeq 10^3M_\odot$ at
redshifts $1+z\sim 16\,\&\,23$, correspondingly. Formation of
less massive starlike subclouds is also hampered. This means
that formation of halos with the virial masses $M\leq 10^6
M_\odot$ at redshifts $z_f\leq 2.5$ is not usually accompanied
by a noticeable star formation.

Other example -- evolution of two halos with masses $M_{vir}=
10^9 M_\odot$ and $M_{vir}=0.3\cdot 10^9M_\odot$ formed at
$z_f\sim 1$ is presented in Fig. \ref{temp191}. In this case
the gravitationally bounded starlike subclouds with $M_{scl}
\simeq M_J\approx 10^3M_\odot$ can be formed at $1+z\leq
9.8\,\&\,7.5$, correspondingly. This means that at $z_f\leq 1$
even massive stars can be formed presumably within metal free
halos with  $M_{vir}\geq 10^9M_\odot$.

\subsection{Constrains on the star formation process}

It can be expected that the first stars are formed at $z_{cr}\sim
20 - 30$ within rare DM halos with the low relic concentration
of electrons $x_e\sim 10^{-4}$ and relic entropy of baryons
(\ref{rel}). Both these values are determined at $z\sim 100 -
300$ after the period of hydrogen recombination. As was shown
above with such low concentration of electrons it is not possible
to efficiently form $H_2$ molecules and to cool the gaseous
component of low mass halos. However these processes are
strongly accelerated when the temperature of the compressed
baryonic component exceeds (conventional) level $T_{gas}=10^4K$.
In this case thermal ionization of hydrogen takes place what
rapidly increases the concentrations of both electrons and
$H_2$ molecules, and accelerates cooling of the gaseous component
and formation of starlike clouds.

This means that the condition (\ref{limit}) which restricts this
temperature can be used as an approximate demarcation line on
the plane $M_{vir},z_{cr}$ between regions of rapid and slow star
formation. The more convenient presentation of this line is
\be
M_{vir}\simeq 10^6M_\odot \left[\frac{17.1}{1+z_{cr}}\right]^{10}
\left[\frac{T_{bc}}{10^4K}\right]^2\frac{S_0}{S_{rel}}
\Theta_p^{-3/4}\,.
\label{zmin}
\ee

As is seen from (\ref{zmin}) for each mass of halo there is the
minimal redshift of the halo formation, $z_{cr}$, for which the
baryonic component is rapidly cooled and can form starlike
objects. The halos with the primordial chemical composition that
were formed at redshifts less then some minimal redshift, $z_{cr}
\leq z_{min}(M_{vir})$, practically cannot produce Pop. III stars
and ionizing photons. For example, as is seen from Fig.
\ref{temp310} for the halo with $M_{vir}=0.5\cdot 10^6M_\odot$
formed at $z_{cr}\leq 25$ temperature of the compressed gas
decreases very slowly and the star formation is strongly delayed.
Similar results are presented in Fig. \ref{temp191} for halos
with the relic composition and masses $M_{vir}\leq 0.4\cdot
10^9 M_\odot$ formed at $z_{cr}\leq 10$.

 However it is necessary to note the approximate character of
the relation (\ref{zmin}) and its complex links with
the process of star formation and the masses of formed stars.
More detailed analysis shows that the restriction (\ref{zmin})
is eroded owing to the radial variations of temperature, possible
ionization by the UV background, random variations of the initial
perturbations and so on. This means that stars can be
efficiently formed also in DM halos that are found is some strip
around the line (\ref{zmin}).

These results lead to important conclusions regarding the sources
of the UV photons that caused the reionization of the Universe
at redshifts $z\sim 10 - 11$ (Komatsu et al. 2011; Larson et al.
2011). As is seen from Eq. (\ref{zmin}) at redshifts $z\simeq 10$
the halos with virial masses $M_9\geq 1- 10$ provide most of
Population III stars with $M_{str}\sim 10^2 - 10^3M_\odot$ while
the contribution of less massive halos can be moderate.
The contribution of low mass galaxies to the UV background is also
suppressed by the feedback of supernova explosions (see, e.g.,
Wyithe et al. 2013; Salvadori et al. 2013; Ceverino et al. 2013).

\subsection{Impact of the Lyman - Werner radiation}

The constrain (\ref{zmin}) is strongly enhanced
when the disruption of $H_2$ molecules by the Lyman-Werner (LW)
or $H^-$  ions by the infrared (IR) photons becomes noticeable
(see, e.g. discussion in Loeb \& Barkana, 2001; Mu\~noz et al.,
2009; Wolkott-Green \& Haiman, 2012 and references therein).
These photons are produced by thermal sources of radiation
(such as stars) together with $Ly-c$ and more energetic photons.
As is well known, for the strong reionization at $z\sim 10$ it
is necessary to produce at least one $Ly-c$ photon per baryon.
The allowance for the complex spectral distribution of the
generated UV photons, ionization of He and the heating and
recombination of the IGM can increase this estimate by a factor
of 2 --3. This is the minimal value and in some papers (see, e.g.
Dijkstra, Haiman, Loeb 2004; Madau 2007) production of extra (up
to 10) UV photons per baryon is discussed. But at the same time
the thermal sources produce comparable number of LW photons with
the density $n_{LW}\sim n_b\sim 3\cdot 10^{-7}z^3cm^{-3}$. Such
flux of LW photons practically brings to a halt formation of
$H_2$ molecules and first star (see, e.g., Safranek-Shrader et
al., 2012).

However, generation of the LW photons is strongly suppressed
when the ionization of the hydrogen and helium is provided by
non thermal sources of the UV radiation. Such non thermal
radiation is inevitably generated by matter accreted onto black
holes created by explosions of massive and supermassive stars.
But in contrast with the thermal sources the non thermal sources
do not produce immediately the LW photons and do not decelerate
the process of formation of first stars. At high redshifts the
heating of the intergalactic gas by the soft X-ray background
is not efficient owing to the cooling of ionized baryons by the
inverse Compton scattering, free -- free emission, excitations
of the neutral hydrogen and so on. In this case we can expect the 
moderate increase of the Jeans mass up to
\[
M_J\approx 4\cdot 10^7 T_4^{3/2}z_f^{3/2} M_\odot\,,
\]
and corresponding increase of masses of forming galaxies.
However in this case the more efficient generation of the IR
and hard UV backgrounds can be expected what leads to partial
ionization of HeI and HeII. Perhaps this inference can be
confirmed by observations of tracks of He lines such as
$\lambda=304 A$ and $\lambda=584 A$ shifted by redshift
$1+z_{cr}\geq 10$ to the region of visible spectrum.

Random spatial distribution of the first galaxies and random
variations of generated UV radiation implies that realistic
representation and analysis of the reionization process is
possible only with the representative numerical simulations
that consider at the same time both the process of star
formation and generation of the UV and LW backgrouds. It can be
expected that these processes are separated in space what
requires a representative simulated volume together with a
high resolution.

\section{Scaling relations for the DM dominated objects.}

Simulations show that characteristics of the virialized DM
halos are much more stable than the characteristics of baryonic
component and after formation at $z=z_{cr}$ of virialized DM
halos with $\langle\rho_{vir}\rangle\approx 200\langle\rho(
z_{cr})\rangle$ slow matter accretion only moderately changes
their characteristics (see, e.g., Diemer et al. 2013). Because
of this, we can observe earlier formed high density galaxies
with moderate masses even within later formed more massive but
less dense clusters of galaxies, filaments and other structure
elements. This means that using the model presented in Sec. 2
for description of the observed dSph galaxies and clusters of
galaxies dominated by DM component we can find one--to--one
correspondence between their observed parameters and the so
called redshift of object formation, $z_{cr}$. Of course
according to the Press -- Schechter approach (Press, Schechter,
1974; Peebles 1974) these redshifts characterize the power
spectrum of the density perturbation rather than the real
period of the object formation.

Following this approach to describe the halos formation we use
the function $B(z_{cr})$ (\ref{bbz}) rather than the redshift
$z_{cr}$. For large $z_{cr}\gg 1$  the function $B(z_{cr})$ is
equivalent to the redshift and we can use redshift $z_{cr}$ for
discussion of properties of the dSph galaxies. However many
observed clusters of galaxies are situated at redshifts $z\leq 1$
and in this case these differences become significant.

The simplest way to estimate the redshift of halo formation
is to use Eq. (\ref{nfw-vir}) which can be rewritten as
\[
(1+z_{cr})^3=3M_{vir}/4\pi R_{vir}^3/\Delta_v\langle\rho_m(
z=0)\rangle\,.
\]
However this approach has to deal with unstable ragged periphery
of halos and to reasonably estimate the virial radius it is
necessary to use a complex model dependent procedure.

More stable but more complex way is to use the expression
(\ref{nfw-c}) (or its equivalent) for the matter concentration
and/or parameters of the central core - $r_s\,\&\,\rho_c$.
However in this case it is not possible to achieve high precision
because of the possible impact of baryonic component, complex
shape of these relations and complex procedures of measurement
of these parameters. Non the less this approach  has the largest
potential to analyze the available observations.

Today the parameters of the virialized DM halos are known for
some population of dSph galaxies (Walker et. al., 2009, 2012;
Tollerud et al. 2012) and for many clusters of galaxies (see
e.g. Piffaretti et al. 2011; Kravtsov, Borgani 2012; McDonald 
2013; and references below). Here we present some results of 
such analysis.

Perhaps this approach can be applied also for virialized groups
of galaxies and for galaxies with measured rotation
curves at large distances. However measurement of the virial
radius and the mean density for such objects is problematic
and more indirect approaches must be used for such analysis.

\subsection{Redshift of formation of the dSph galaxies }

Samples of the dSph and And (companions of the Andromeda galaxy)
galaxies include objects in a wide
range of masses, $0.1\leq M_6= M_{gal}/10^6M_\odot\leq 100$,
what allows us to reveal more reliably the mass dependence of
their redshift of formation. Some observed properties of 28
dwarf DM dominated galaxies are compiled in Walker et al.
(2009) and for 13 And galaxies are presented in Table 4 in
Tollerud et al. (2012). In this case we have to deal with
parameters of the central regions at the projected half--light
radius. Moreover, the presented data are recalculated from
actual observations (Walker et al. 2009; Walker 2012) and, so,
their reliability is limited and scatter is large. In spite of
this it is interesting to compare characteristics of these
galaxies and observations of clusters of galaxies with the
theoretical expectations of Sec. 3.

First of all for the sample of 28 dSph galaxies we can roughly
estimate the dimensionless size of the region under consideration.
To do this we compare the observed masses, radii and velocity
dispersions with expectations (\ref{tnfw}). For this sample we get
\[
\langle\sqrt{r_{ob}/r_s}\rangle\approx \left\langle\frac{G
M_{ob}}{2r_{ob}\sigma^2_v}\right\rangle\approx 1.3\pm 0.02\,.
\]
This means that the observed values of $M_{ob},\,\sigma_v,\&
\,r_{ob}$ are related to the model parameters as follows:
\be
r_{ob}\sim 1.7r_s,\quad M_{ob}\sim 0.36M_{vir},\quad \langle
\rho_{ob}\rangle\sim 0.1\rho_c\,.
\label{real}
\ee
For both the NFW and Burkert models these corrections are quite
similar to each other because they consider regions where
$r_{ob}\geq r_s$.

\begin{figure}
\centering
\epsfxsize=7.2 cm
\epsfbox{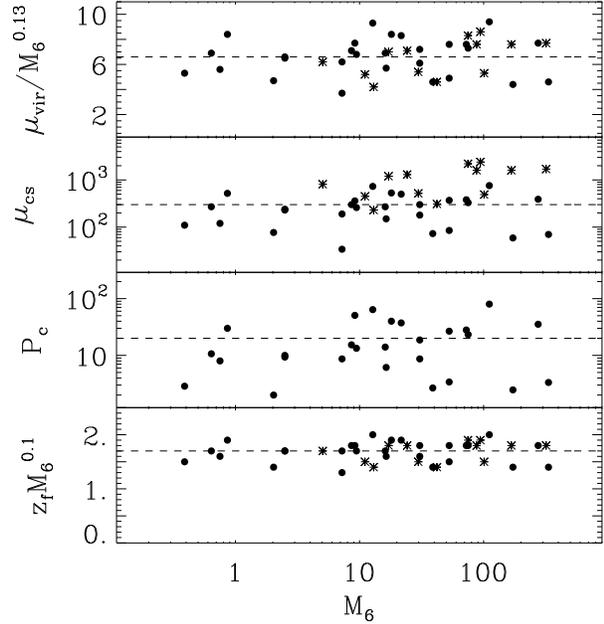}
%%\epsfbox{dsph_16.eps}
\vspace{0.95cm}
\caption{Functions $\mu_{vir}/M_6^{0.13}$, $\mu_{cs}$, $P_c(M_6)$
and $z_fM_6^{0.1}$ are plotted for the observed samples of 28 dSph
galaxies (points) and 13 And galaxies (stars). Fits (\ref{zfm}),
(\ref{pnfw-o}), (\ref{mu_dsph}) and (\ref{svdsph}) are plotted by
dashed lines.
}
\label{zdsph}
\end{figure}

In order to find the redshift of object formation,
$z_f=(1+z_{cr})/10$, we can use two approaches. Firstly
we can use the estimates of the central density (\ref{dnfw})
which can be expressed with the help of (\ref{real}) through
the observed $M_{ob}$ and $\langle\rho_{ob}\rangle$
\be
%%z_f^{10}=
z_{f\rho}^{10}\approx \frac{150}{\sqrt{M_6}}
\langle\rho_{ob}\rangle\frac{pc^3}{M_\odot}\,,
\label{zdens}
\ee
where $M_6=M_{ob}/10^6M_\odot$ is the observed mass of DM
galaxy. The great advantage of this method is a weak dependence
of $z_{fr}$ on  $\rho_{ob}$ what attenuates the impact of errors.
For comparison we can use expression (\ref{size}) for the
typical size of the central regions,
\be
z_{fr}^{10/3}\approx 0.5M_6^{1/6}/r_{kpc}\,,
\label{zsize}
\ee
where $r_{kpc}$ is the observed radius, $r_{ob}$, in kpc.
Both estimates are quite similar to each other and we have
\[
z_f\simeq z_{f\rho}\approx 0.9z_{fr}\,.
\]
The redshifts of formation of galaxies are spread between
values $1+z_{cr}=8$ for $Sgr^c$ and $1+z_{cr}=20$ for Segue 1.
The fit of the mass dependence of $z_{cr}(M)$ is
\[
z_fM_6^{0.1}\approx 1.7(1\pm 0.12)\,,
\]
\be
1+z_{cr}\approx 17(1\pm 0.12)M_6^{-0.1}\approx 3(1\pm 0.12)
M_{13}^{-0.1}\,,
\label{zfm}
\ee
 These results are presented in Fig. \ref{zdsph}.

For 13 And galaxies the masses and half--light radii are listed
in Table 4 in Tollerud et al. (2012). For these objects we can
use the expression (\ref{zsize}) to estimate their redshift of
formation. For this sample we get
\be
z_fM_6^{0.1}\approx 1.6(1\pm 0.11),\quad 1+z_{cr}\approx 16(1\pm
0.11)M_6^{-0.1}\,,
\label{and1}
\ee
what is identical with (\ref{zfm}). The maximal $1+z_{cr}\approx
14$ is obtained for the galaxy And XVI.

These results are also presented in Fig. \ref{zdsph}. However it
is necessary to note that parameters of the galaxies And IX and
And XV presented in both surveys are noticeably different.

Comparison of the relations (\ref{pnfw}) and (\ref{zfm}) shows
that the weak mass dependence can be expected also for the
typical DM pressure, $P_c(M_{vir},z_f)$. Analysis of the sample
of dSph galaxies results in the following estimate
\be
P_c(M_6)\approx 20(1\pm 0.9)eV/cm^3\,.
\label{pnfw-o}
\ee
This function is plotted in Fig. \ref{zdsph}. It is very close 
to demarcation value (\ref{limit}) but is more sensitive to 
(random) scatter of the observed parameters then $z_f$.

Combining Eq. (\ref{zfm}) with Eqs. (\ref{size}) and
(\ref{dnfw}) we can directly link parameters of the core
and virial masses of halos
\be
\rho_c\sqrt{M_{vir}}\approx const,\quad r_s/\sqrt{M_{vir}}
\approx const\,.
\label{rrm}
\ee
However these links are not so strong and for these relations
scatter can be as high as  50 -- 100\% what decreases their 
usefulness.

\subsection{Redshift of formation of clusters of galaxies}

Now there are more or less reliable observational data at least
for $\sim 300$ clusters of galaxies (Ettori et al. 2004;
Pointecouteau et al. 2005; Arnaud et al., 2005; Pratt et al.,
2006; Zhang et al., 2006; Branchesi et al., 2007; Vikhlinin et
al., 2009; Pratt et al. 2010;  Suhada et al. 2012; Moughan et
al. 2012; Babyk et al. 2012; Fo\H{e}x et al. 2013; Bhattacharya
et al. 2013). However, the main cluster characteristics are not
directly observed and are obtained by a rather complex procedure
(see, e.g., Bryan\,\&\,Norman 1998; Vikhlinin et al. 2009; Lloyd--
Davies et al. 2011; McDonald M., et al., 2013).
In particular they relate the virial mass and radius of each
cluster with the critical density of the Universe at the
observed redshift, $\rho_c(z_{obs})$,
\[
M_{vir}=4\pi/3R_{vir}^3 500\rho_c(z_{obs})= 250
R_{vir}^3H^2(z_{obs})/G\,.
\]

In fact this assumption identifies the redshift of cluster
formation with the observed redshift. This assumption is
questionable for majority of clusters as quite similar clusters are
observed in a wide range of redshifts. It distorts all published
cluster characteristics and often makes impossible to use the
published characteristics of cores for further discussions.
The matter concentration is measured with a reasonable precision
only for 25 clusters of sample CLS-25 combined from samples
CLS-10 (Pointecouteau et al. 2005), CLS-12 (Vikhlinin et al.
2006) and CLS-18 (Bhattacharya et al. 2013). The central
regions of many clusters are influenced by cooling baryonic
component (see, e.g., Pratt et al., 2009, 2010) but for these
three samples the concentrations are determined with precision
of $\sim 10 - 15\%$ what allows us to estimate the redshift of
formation, $z_{cr}$, and both the central and the virial surface
densitis of clusters, $\mu_{cs} =r_s\rho_c$ and $\mu_{vir}=
R_{vir}\langle\rho_{vir}\rangle$.

For this sample the redshift $z_{cr}$ can be obtained from the
relation (Dolag 2004):
\be
1+z_{cr}\approx \frac{11}{c(M_{vir},z_{cr})M_{13}^{0.1}},\quad
M_{13}=M_{vir}/10^{13}M_\odot\,,
\label{c-cls}
\ee
\[
\langle 1+z_{obs}\rangle =1.09(1\pm 0.05),\quad \langle
c\rangle\approx 4.05\pm 0.9\,,
\]
\be
\langle 1+z_{cr}\rangle\approx 2.1(1\pm 0.24)\,.
\label{cls10}
\ee
The maximal values $1+z_{cr}\approx 3.6$ and $1+z_{cr}\approx
2.6$ are obtained for the cluster $MKW4$ with $M_{13}\approx
7.7$ and for the cluster $A262$ with $M_{13}\approx 8.3$
observed at $z_{obs}=0.02$ and $z_{obs}=0.016$ (Vikhlinin et
al. 2006; Bhattacharya et al. 2013). For these clusters $z_{cr}
\gg z_{obs}$ what confirms differences between the redshift of
cluster formation and random observed redshift at least for
clusters with $z_{obs} \ll 1$.

However as was noted above for redshifts $z_{cr}\leq 1$ the
cluster formation is determined by the function $B(z)$
(\ref{bbz}) rather than by the redshift $z_{cr}$. Thus for
these samples we get
\be
\langle B^{-1}(z_{cr})\rangle\approx 1.63(1\pm 0.2)=
2.3(1\pm 0.2)M_{13}^{-0.1}\,.
\label{b20}
\ee
This value is quite comparable with the estimates (\ref{zfm}) 
which can be rewritten for dSph galaxies as follows
\be \langle
B^{-1}(z_{cr})\rangle\approx 2.22(1\pm 0.12)M_{13}^{-0.1}\,.
\label{Bdsph}
\ee
However large scatter and uncertainties in both estimates
(\ref{zfm}) and (\ref{b20}) prevent more detailed comparison of
these results.

\begin{figure}
\centering
\epsfxsize=7. cm
%%\epsfbox{cls_83bzm.eps}
\epsfbox{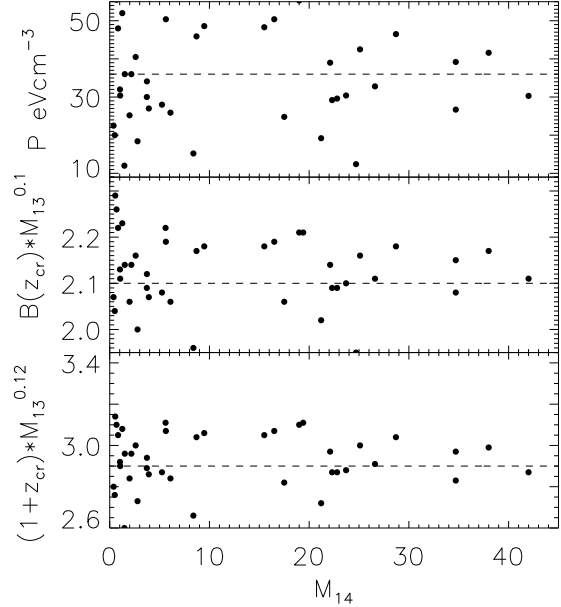}
\vspace{0.95cm}
\caption{For 44 clusters from the sample CLS-83 the redshift 
formation, $(1+z_{cr})M_{13}^{0.12}$ and $B(z_{cr})M_{13}^{0.1}$, 
and of the central pressure, $P_c$, are plotted vs. mass of clusters. 
Dashed lines show fits (\ref{hot-cls}).
}
\label{cls_83sm}
\end{figure}

The more interesting sample CLS-83 (McDonald et al. 2013) 
contains parameters of central regions of 83 clusters with
redshifts $z\geq 0.3$, namely, baryonic density and temperature, 
$n_c\,\&\,T_c$. Using relations (\ref{shock}) and (\ref{nfw-ent}) 
we can estimate for these clusters the redshift of formation 
$z_{cr}$ and mass $M_{13}$ 
\be
M_{13}\approx 10^{-4}\left(\frac{44cm^{-3}}{n_c}\right)^{1/2}
\left(\frac{T_c}{9.5eV}\right)^{3/2}f_m(c)\,,
\label{cls83}
\ee
\[
1+z_{cr}\approx 10.\left(\frac{n_c}{44cm^{-3}}\right)^{1/8}
\left(\frac{9.5eV}{T_c}\right)^{3/40}\epsilon^{-0.3}\,.
\]
Ror 44 clusters of this sample with the central pressure 
$P_c=n_cT_c\leq 70~eVcm^{-3}$ we get 
\be
\langle P_c\rangle\approx 36(1\pm 0.37) eV cm^{-3}\,,
\label{pcls}
\ee
\be
\langle (1+z_{cr})\rangle\approx 2.9(1\pm 0.05)M_{13}^{-0.12},
\quad 0.3\leq z_{cr}\leq 1.5\,,
\label{hot-cls}
\ee
\[
\langle B(z_{cr})\rangle\approx 2.1(1\pm 0.04) M_{13}^{-0.095}\,.
\]
These results are plotted in Fig. \ref{cls_83sm}. 

It is important that in spite of the large difference in masses 
of these clusters $(M/M_\odot\geq 10^{13})$ and dSph galaxies with 
$(M/M_\odot\leq 10^{9})$  estimates (\ref{pcls}) are very close 
to (\ref{pnfw-o}) while (\ref{hot-cls}) are quite similar to 
(\ref{cls10})--(\ref{Bdsph}). For other 39 clusters of this sample 
all characteristics are strongly distorted owing to cooling of 
gaseous component what significantly increases scatter of the 
final estimates.

\subsection{Surface density of DM dominated objects}

In set of publications (e.g., Spano et al., (2008); Donato et
al. 2009; Gentile et al. 2009; Salucci et al., 2011) it is
found that the central surface density, $\mu_{cs}=r_s \rho_c$,
is almost the same across a wide range of galaxies of different
types and luminosities. Thus for 36 spiral galaxies Spano et al.
(2008) obtains
\[
\langle\mu_{cs}\rangle\approx 150^{+100}_{-70}M_\odot pc^{-2}\,.
\]
For the dSph galaxies Donato et al. (2009) estimates this
surface density as
\[
\langle\mu_{cs}\rangle\sim 140^{+80}_{-30}M_\odot pc^{-2}\,,
\]
while Salucci et al. (2011) infer that
\[
\log(\rho_c)= -\alpha\log(r_s),\quad 0.9\leq\alpha\leq 1.1\,.
\]
Here we will check these inferences with the available data sets.

\subsubsection{Surface density of the dSph galaxies}

In Sec. 4.1 links between the core and virial parameters of the
dSph galaxies and their redshift of formation were discussed. The
weak mass dependence of the observed DM surface density of the
same dSph objects is other demonstration of the same links.
Indeed, as is seen from (\ref{surf}) the weak mass dependence
of the function
\be
\langle\mu_{cs}\rangle=r_s\rho_c\approx F_{srf}^{2/3}\Theta_{sc}
M_\odot/pc^2,\quad \Theta_{sc}=\delta_r\epsilon^3/f_m^{3/2}(c)\,,
\label{mufc}
\ee
follows immediately from the weak mass dependence of the function
\be
F_{srf}=M_6z_f^{10}\simeq const.,
\label{fms}
\ee
which determines also the weak mass dependence of the redshift of
formation $z_{cr}$ discussed in Sec. 4.1\,. Here $c(M_9, z_{cr})$
is the concentration (\ref{nfw-c}), and the virial dimensionless
mass of objects $f_m(c)$ is given by (\ref{fmass}) for the NFW
density profiles.

Using the estimates (\ref{zfm}) we get
\[
\langle\mu_{cs}\rangle\approx 230M_\odot/pc^2\,,
\]
while the direct estimates with the full sample of 28 dSph
galaxies give
\be
\langle\mu_{cs}\rangle\approx 300(1\pm
0.66\pm 0.51)M_\odot/pc^2\,.
\label{mu_dsph}
\ee
Here the first uncertainty is connected with the scatter of
$\mu_{cs}$ over the sample, while the second one characterizes
the precision of separate measurements. These results are
presented in Fig. \ref{zdsph}.

Large scatter of the surface density strongly decreases its
significance and possible applications. Non the less for
galactic scales the weak mass dependence of the surface
density is confirmed by the weak mass dependence of the
redshift of formation.

In accordance with (\ref{vsurf}) the virial surface density
of dSph galaxies is weakly dependent on its virial mass and
is described by the relation
\be
\langle\mu_{vir}\rangle\approx 6.6M_6^{0.13}(1\pm 0.21)
M_\odot/pc^2\,,
\label{svdsph}
\ee
where as before $M_6=M_{vir}/10^6M_\odot$ and $M_{vir}\approx
M_{half}/0.36$. This result is consistent with (\ref{zfm}).
The function $\mu_{vir}/M_6^{0.13}$ is plotted in Fig.
\ref{zdsph}.

\subsubsection{DM surface density for clusters of galaxies}

It is necessary to remind that if for galaxies  we had
$\epsilon(M,z_{cr})\sim 1$ in relations  (\ref{nfw-c}) -
(\ref{pnfw}) then for some clusters of galaxies $\epsilon\gg 1$
and we get for their DM surface density
\be
\mu_{cs}\approx 204\left[\frac{M_{13}}{(1+z_{cr})^4}\right]^{1/6}
 \Theta^*\frac{M_\odot}{pc^2},\quad \Theta^*=
\frac{\delta_r^{2/3}\varepsilon^2}{f_m(c)}\,,
\label{mucls}
\ee
\[
\varepsilon=1+3.7\cdot 10^{-2}M_{13}^{1/4}(1+z_{cr})^4,
\,\,M_{13}=M_{vir}/10^{13}M_\odot\,.
\]
This means that relations (\ref{mufc}, \ref{fms}) which are
valid for galaxies cannot be applied for massive clusters
of galaxies.

Using the expression (\ref{mucls}) with $\langle 1+z_{cr}\rangle$
estimated by (\ref{cls10}) we get
\be
\mu_{cs}\approx 124M_{13}^{1/6} M_\odot/pc^2\,,
\label{cls-cs}
\ee
what is a rough estimate owing to the large scatter of $z_{cr}$.
More accurate results can be obtained from expressions
(\ref{nfw-vir}):
\be
\mu_{cs}=\frac{M_{vir}}{4\pi r_s^2f_m(c)}=\frac{100}{4\pi}
\frac{M_{13}}{f_m(c)}\frac{c^2}{R^2_{Mpc}}\frac{M_\odot}{pc^2}\,,
\label{mu-cls}
\ee
where as before $c(M,z_{cr})$ is the halo concentration,
$M_{13}= M_{vir}/10^{13}M_\odot$, and $R_{Mpc}$ is the
virial radius of cluster in Mpc. With this relation we get
for the sample CLS-25
\be
\langle\mu_{cs}\rangle\approx 415(1\pm 0.3)
\frac{M_\odot}{pc^2} \approx 150(1\pm 0.25)M_{13}^{0.3}
\frac{M_\odot}{pc^2}\,.
\label{surfcls}
\ee
This estimate differs from that obtained for galaxies
(\ref{mu_dsph}) because it depends on mass. The fact that
(\ref{cls-cs}) and (\ref{surfcls}) are different shows that
these results depend on  the averaging procedure.

The virial surface density of clusters is closely linked
with their central surface density, $\mu_{cs}$ (\ref{mu-cls}).
Thus, for the same sample CLS-25 we get
\be
\langle\mu_{vir}\rangle=3\langle f_m(c)\mu_{cs}/c^2\rangle
\approx 17(1\pm0.28)M_{13}^{0.35}M_\odot/pc^2\,.
\label{vsrcls}
\ee

These results can be compared with recently published data by
Babyk et al. (2012). For the sample CLS-30 of 30 clusters
randomly selected from this survey we get
\be
\langle\mu_{vir}\rangle\approx 13(1\pm 0.11)M_{13}^{0.35}
M_\odot/pc^2\,.
\label{cls-30}
\ee
However in this survey all cluster characteristics are found
with the popular assumption that $z_{cr}=z_{obs}$ what
distorts their virial parameters, $M_{vir}\,\&\,R_{vir}$,
and makes impossible discussion of characteristics of the cluster
cores. Prominent differences between scatters (\ref{vsrcls}) 
and (\ref{cls-30}) are caused mainly by the impact of cluster 
description rather than by their physical properties.

The relatively small interval of observed cluster masses and
the limited reliability and precision of the complex procedure
of reconstruction of cluster characteristics (see, e.g., Bryan\,
\&\,Norman 1998; Pointecouteau et al. 2005; Pratt et al. 2009;
Vikhlinin et al. 2009; Lloyd--Davies et al. 2011) strongly
restricts the applicability of discussed scaling relations.

\section{Conclusions}

Abundant simulations show that the formation of virialized DM
halos is a complex multistep process which begins as the
anisotropic collapse in accordance with the Zel'dovich theory
of gravitational instability (Zel'dovich 1970). During later
stages the evolution of such objects is complicated  and it
goes through the stages of violent relaxation and merging.
Non the less after a period of rapid evolution the main
characteristics of the high density virialized DM halos become
frozen and their properties are slowly changing owing to the
accretion of diffuse matter and/or the evolution of their
baryonic component.

The basic properties of the relaxed DM halos are determined
by their global characteristics, namely,  their mass, angular
momentum and entropy generated in the course of violent
relaxation of compressed matter. Basically the structure of
such halos is similar to the structure of clusters of
galaxies (see, e.g., Tasitsiomi et al. 2004; Nagai et al.
2007; Croston et al. 2008; Pratt et al., 2009, 2010; Arnaud
et al. 2010; Kravtsov \& Borgani 2012). In particular in
these papers the generalized NFW model proposed in Zhao
(1996) and Nagai et al. (2007) is discussed.

The basic properties of halos can be reproduced in the
framework of the popular spherical model of halos
formation. Such models have been discussed for many years
(see, e.g., Peebles 1967; Zel'dovich \& Novikov 1983; Fillmore
\& Goldreich 1984; Bryan \& Norman 1998; Lithwick, Dalal 2011).
However this model ignores many important features of the
process of halos formation and is based on the assumption
that during a short period of the spherical collapse at
$z\approx z_{cr}$ the DM forms virialized halos with
parameters which later vary slowly owing to the successive
matter accretion (see, e.g., discussion in Bullock et al.
2001; Diemer et al. 2013).

In this paper we use the analytical description of the
virialized spherical DM halo with the NFW density profile
proposed in Klypin et al. (2011). Such approach allows us
to formulate in Sec. 2 two parametric spherical model of
virialized halos which is specified by the virial halo mass,
$M_{vir}$, and redshift of formation $z_{cr}$. Of course
this redshift is only some conventional characteristic of
the mean density of virialized halos or other corresponding
parameters. However, it can be used in order to roughly
characterize the period of halos virialization what in
turn allows to order the observed halos with respect to
the (conventional) moment of formation. It also opens up
the comparatively simple way to reveal correlations of thus
introduced redshifts with the shape of the initial power
spectrum.

\subsection{Mass dependence of the redshift of formation of
DM halos}

These problems are discussed in Sec. 4 where we obtained the
approximate relation between the virial mass of DM objects
and their redshifts of formation. According to the commonly
accepted hierarchical model of galaxy formation at high redshifts
the formation of low mass galaxies dominates and the typical mass
of formed galaxies successively increases with time. These
expectations are illustrated by the expressions (\ref{zfm}),
(\ref{cls10}) -- (\ref{Bdsph}). The high redshifts of formation
of dSph and And galaxies correlate well with their low metal
abundance and help us to reconstruct the history of the Local
Group discussed for instance by Peebles (1995, 1996), Klypin et
al. (2002, 2003).

Combining the resulting estimates (\ref{b20}) and (\ref{Bdsph})
for mass dependence of the redshift of formation for both DM
dominated observed dSph galaxies and clusters of galaxies we
conclude that for these objects
\be
\langle B^{-1}(z_{cr})\rangle\approx 2.3(1\pm 0.15)/M_{13}^\beta\,,
\label{mz}
\ee
\[
\beta\sim 0.1,\quad 10^5M_\odot\leq M_{vir}\leq 10^{14}M_\odot\,.
\]

With respect to the general theory of gravitational instability
for the DM objects the expression (\ref{mz}) quantify the mass
dependence of the mean redshift of formation, $z_{cr}(M_{13})$,
or correlation between these redshifts and the shape of the
initial power spectrum of perturbations. Indeed according to
the standard $\Lambda$CDM cosmology low mass objects (such as
galaxies) are presumably formed earlier than more massive
galaxies and clusters of galaxies and the expression (\ref{mz})
illustrate this statistical tendency.

Thus for the standard CDM -- like power spectrum $p(k)$ (Bardeen
et al. 1986) we have for the density perturbations (Klypin et al.
2011)
\be
\sigma_m^2=4\pi\int_0^\infty p(k)W^2(kR)k^2dk,\quad
\sigma_m\propto M^{-0.1}\,,
\label{sig_m}
\ee
for $10kpc\leq R\leq 10Mpc$,\,\,$10^5\leq M/M_\odot\leq 10^{14}$.
Here $W(kr)$ is the standard top--hat window function. Following
the Press -- Schechter approach (Press\,\&\,Schechter 1974;
Peebles 1974; Mantz et al. 2010) we can determine the redshift
of objects formation from the condition
\[
B(z_{cr})\,\sigma_m(M)\approx const,\quad B^{-1}(z_{cr})\propto
\sigma_m(M)\propto M^{-0.1}\,.
\]
This result is consistent with (\ref{mz}) and confirms that the
CDM--like power spectrum can be extended at least down to the
scale of $\sim 10kpc$. However this approach does not allow us
to obtain an independent estimate of the small scale amplitude
of perturbations. More detailed comparison of the mass
dependence of the redshift of formation of galaxies and clusters of
galaxies requires much more precise estimates of observational
parameters of both galaxies and clusters of galaxies.

For completeness it is interesting also to consider objects with
intermediate masses $M_{vir}\sim 10^{10}M_\odot - 10^{12}M_\odot$.
The virialized groups of galaxies and the far periphery of
isolated galaxies can be used for such analysis if it is possible
to confirm that they are dominated by DM and to estimate
their virial characteristics. Perhaps it is simpler to estimate
the correlation between the measured circular velocity $v_c^2=
GM_{vir}/R_{vir}$ and the virial radius $R_{vir}$. For $(1+z_{cr})
\propto M^{-\beta}$ (\ref{mz}) we can expect that
\be
v_c^2=GM_{vir}/R_{vir}\propto R_{vir}^\gamma,\quad
\gamma=(2-3\beta)/2(1+\beta)\,,
\label{vcr}
\ee
and $\gamma=0.8$ for $\beta\approx 0.1$.

\subsection{Formation of the first  galaxies}

It is quite interesting to compare the limits (\ref{zmin}) with
the observed properties of low mass galaxies (\ref{zfm}\,\&\,
\ref{and1}). The first one restricts the virial masses of DM halos
allowing for the rapid creation of metal free stars and the UV
radiation. The second one considers the most probable masses
of DM halos forming at the same redshifts as is suggested by
observations of the dSph galaxies.

\begin{figure}
\centering
\epsfxsize=7. cm
%%\epsfbox{dsph_8.eps}
\epsfbox{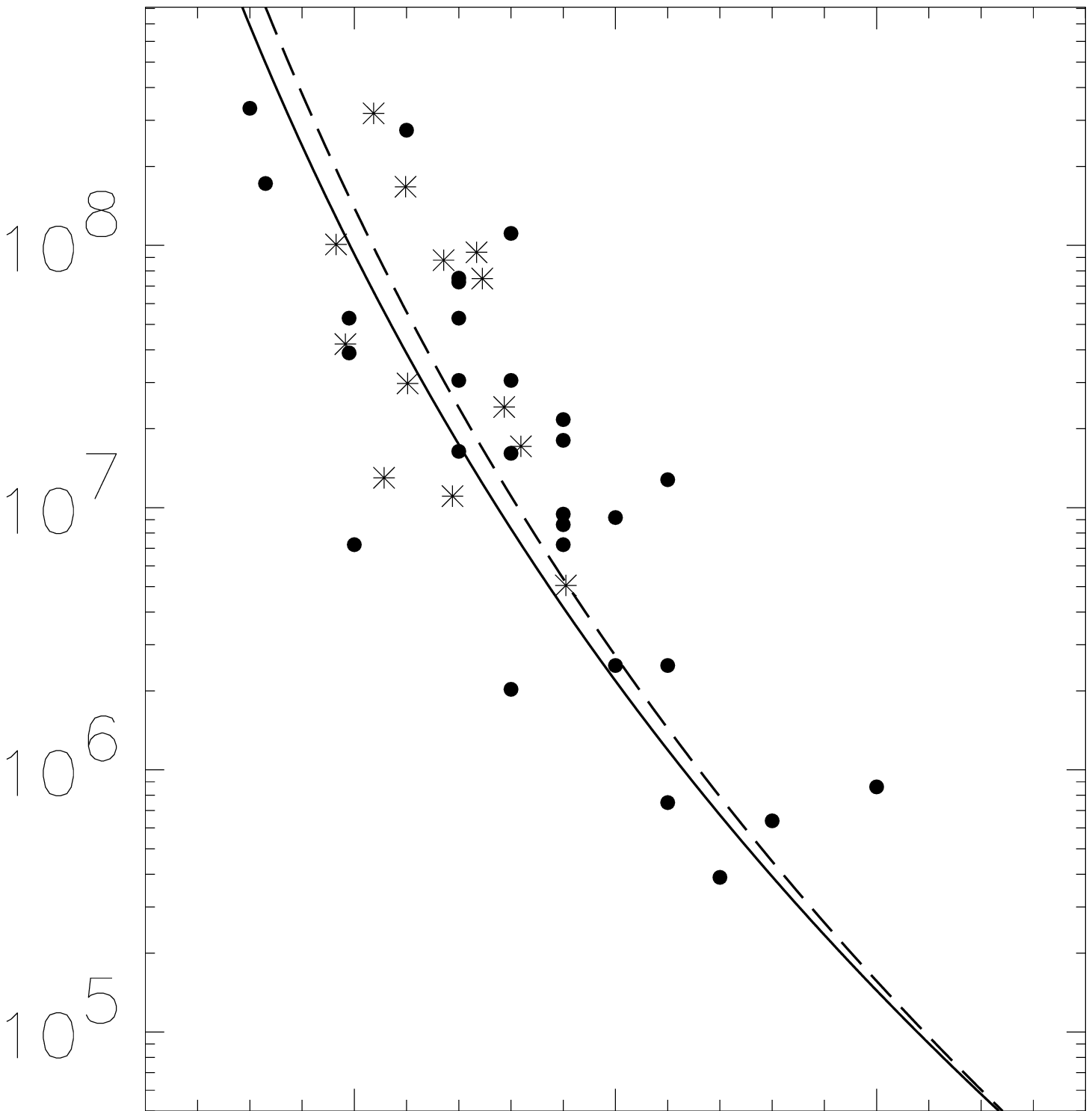}
\vspace{0.95cm}
\caption{The observed masses of dSph (points) and And (stars) 
galaxies, $M_{vir}/M_\odot$, are plotted vs. their redshifts 
of formation, $z_{cr}$. Dashed line shows the fit (\ref{f2}). 
Solid line shows the redshift dependence of the minimal virial 
mass of DM halos with rapid formation of first stars (\ref{zmin}, 
\ref{f1}).
}
\label{mzf-bet}
\end{figure}

Such comparison is presented in Fig. \ref{mzf-bet} where the
expected minimal masses of the DM halos with rapid star formation
(Eqs. (\ref{zmin}) and (\ref{f1}))
 \be
M_{vir}=[17.1/(1+z_{cr})]^{10}\,10^6M_\odot\,,
\label{f1}
\ee
are compared with the observed masses and redshift of formation
for the dSph galaxies (points, stars and fit (\ref{f2})):
\be
M_{gal}=[17.6/(1+z_{cr})]^{10.5}\,10^6M_\odot\,.
\label{f2}
\ee

\begin{figure}
\centering
\epsfxsize=7. cm
%%\epsfbox{cls_83hr.eps}
\epsfbox{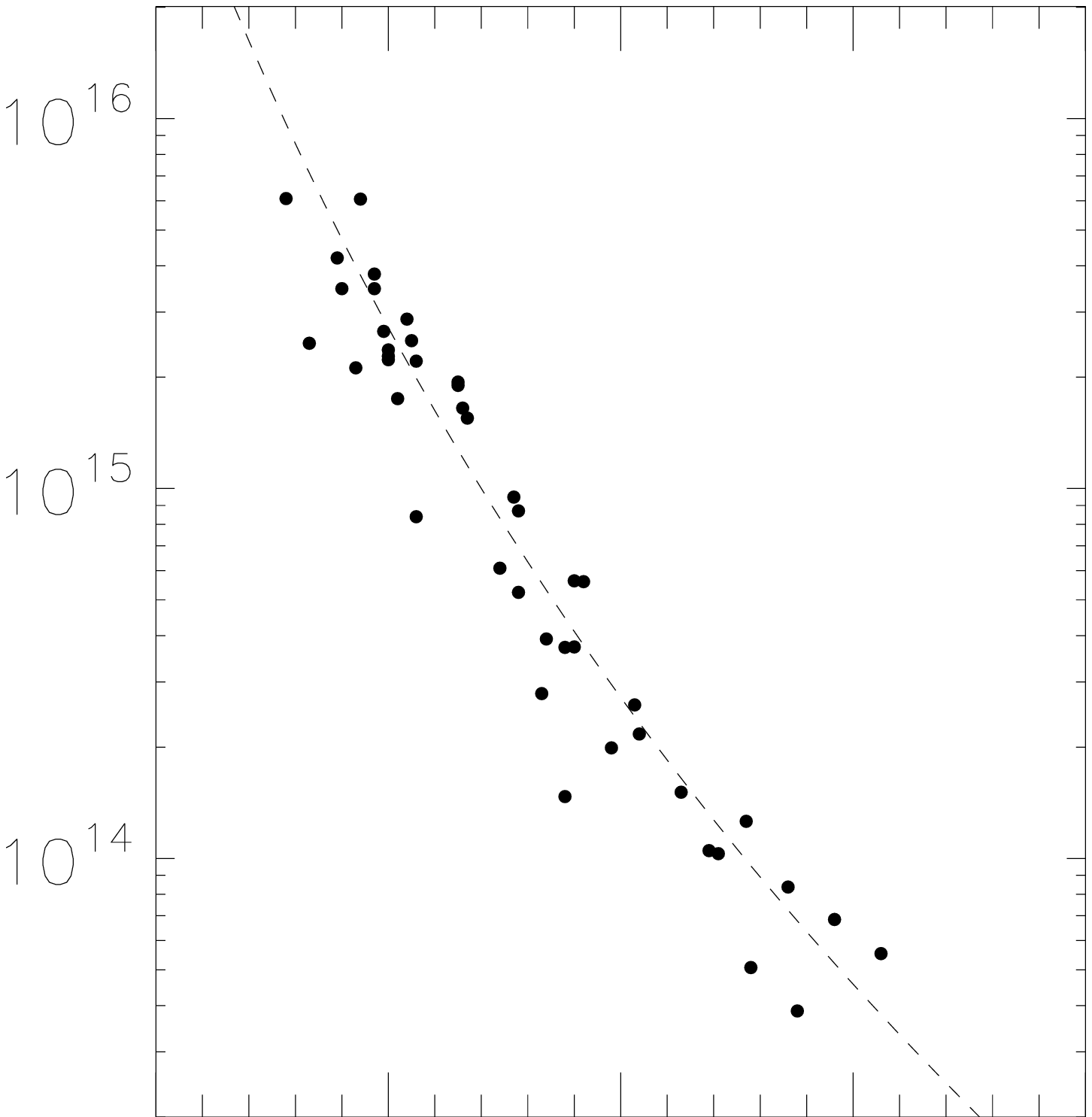}
\vspace{0.95cm}
\caption{For 44 clusters of the sample CLS-83 with $P_c=n_cT_c\leq 
70eV/cm^3$ the virial halo masses, $M_{vir}/M_\odot$, are plotted 
vs. their redshift formation, $1+z_{cr}$. Dashed line shows the 
fit (\ref{f83}).
}
\label{mzf83}
\end{figure}

For 44 clusters of sample CLS-83 with $P_c\leq\,70eV/cm^3$ 
the correlation of the virial mass of halos and their redshift of 
formation is fitted by expression
\be
M_{cls}=[17/(1+z_{cr})]^810^{6}M_\odot\,,
\label{f83}
\ee
and is plotted in Fig. \ref{mzf83}. As is seen from 
(\ref{cls83}) the cooling of baryonic component artificially 
decreases the estimate of virial mass $M_{vir}$ and increases the 
estimate of redshift $z_{cr}$ what enchances the scatter of 
points in Fig. \ref{mzf83}. In spite of this the similarity of 
expressions (\ref{f1}) - (\ref{f83}) reflects the close link of 
all these objects formed with the joint power spectrum of initial 
perturbations. Weaker variation $M_{cls}$ as a function of $z_{cr}$ 
(\ref{f83}) as compared with (\ref{f2}) is naturally explained by 
weaker mass dependence of the amplitude $\sigma_m(M)$ for cluster 
masses.

The complex process of retrieval of considered characteristics of 
the dSph galaxies (see, e.g., Walker 2012) and the plausible impact 
of their prolonged evolution -- such as the probable tidal striping
-- makes detailed discussion of the observed properties of such 
objects unreliable. In spite of this the comparison performed in 
Fig. \ref{mzf-bet} is interesting. First of all it confirms the
probable formation of metal free galaxies with $M_{vir}\leq 10^9
M_\odot$ at redshifts $z\sim 20 - 8$ what agrees well with both
other observations (see, e.g., Wyithe et al. 2013) and theoretical
expectations discussed in Sec. 5.1. 

On the other hand
this Figure shows that the dSph galaxies are concentrated near
the joint approximate boundary (\ref{zmin}, \ref{zfm}). Such
concentration indicates that the observed dSph galaxies can be
really related with the earlier DM objects with various rate of
star formation. Thus in objects disposed to the right of the
demarcation lines (\ref{zmin}, \ref{zfm}) the rapid star
formation can be expected. In contrast in objects disposed to
the left of these lines the star formation can be partly
inhibited either by both the low ionization of the compressed matter
and by the impact of LW background. It can be later stimulated
by an action of external factors such as, for example, ionizing
UV radiation of external sources after dissipation of the LW
background. 

It is interesting also to compare other observed
properties of these groups of dSph galaxies such as their
metallicity and the possible populations of black holes.

These results show also that the most efficient star formation
takes place in halos with $10^7\leq M/M_\odot\leq 10^9$
(\ref{zmin}) at redshifts $z\sim 13- 8$ when probably the
reionization actually occurs. This inference is consistent
with observations of galaxies at redshifts $z\geq 7$ with
$M\sim 3\cdot 10^8 - 10^9M_\odot$ (Ouchi et al. 2009; Schaerer
\& de Barros 2010; Gonzales et al. 2010, 2012; Oesch et al.
2013; Ellis et al. 2013). However, observations of Bouwens et
al. (2011, 2012) indicate a possible more significant
contribution of less massive galaxies.

All theoretical expectations can be essentially corrected by
possible impact of the UV, LW and/or IR background (see, e.g.
discussion in  Loeb \& Barkana 2001; Mu\~noz et al. 2009;
Wolkott-Green \& Haiman 2012 and references therein). As was
noted in Sec. 3.5 the production of the UV background required
for reionization by stars or other sources of radiation with
the thermal spectrum is inevitably accompanied by formation
of the corresponding LW background and deceleration of the
process of star formation. In this case the UV radiation
generated by matter accretion onto black holes with various
masses can become dominant and can really determine the
reionization. This verifies that such non thermal sources
of the UV radiation can be considered as very promising ones
and can be actually responsible for the reionization (see,
e.g., Madau \,\&\,Rees 2001; Reed 2005; Meiksin 2005,\,2009;
Madau 2007; Giallongo et al. 2012). Perhaps, the contribution
of such sources can be confirmed by observations of tracks of
He lines such as 304 A and 584 A shifted to the redshift of
reionization.

\subsection{The DM surface density of relaxed objects}

If the redshift of formation of DM dominated objects can be
directly linked with the power spectrum of density perturbations
then both the central and virial surface densities of these
objects, $\mu_{cs}\,\&\,\mu_{vir}$, should depend upon the
processes of violent relaxation of compressed matter and
characterize this process. In particular, as it is seen from
relations (\ref{surf}, \ref{mucls}), we can expect a weak
mass dependence of the central surface density $\mu_{cs}(
M_{vir})$ at galactic scale (\ref{mu_dsph}) but these
expectations are distorted at clusters of galaxies scale
(\ref{surfcls}). The virial surface density only weakly
dependence on mass at both galactic and clusters of
galaxies scales.

Both surface densities are closely linked with the DM density
profile formed in the course of violent relaxation of the
compressed matter and are determined by the action of the same
factors. Non the less it can be expected that a weak mass
dependence of the central surface density, $\mu_{cs}$, can be
observed across wide set of objects of galactic scales what
will be an important additional evidence in favor of the
standard shape of small scale initial power spectrum.

\subsection*{Acknowledgments}
This paper was supported in part by the Russian Found of
Fundamental Investigations grants Nr. 11-02-00244, and
NS-2915.2012.2 and by the Polish Ministry of Science and
Higher Education grant NN202-091839. AD thanks S.Pilipenko,
B.Komberg and M. Sharina for useful comments. We wish to
thank the anonymous referee for many valuable comments.

\end{document}